\documentclass[11pt]{article}

\usepackage{amssymb}
\usepackage{amsthm}
\usepackage{amsmath}

\usepackage{graphicx}

\newtheorem{theorem}{Theorem}[section]

\newtheorem{remark}[theorem]{Remark}

\newcommand{\D}{{\mathrm{d}}}
\begin{document}
\title{Jordan form, parabolicity and other features of change of type transition for hydrodynamic type systems.}
\author{B.G. Konopelchenko \\
Dipartimento di Matematica e Fisica ``Ennio de Giorgi", Universit\`{a} del
Salento \\
INFN, Sezione di Lecce, 73100 Lecce, Italy \\
{konopel@le.infn.it} \\
\mbox{} \\
G. Ortenzi \\
Dipartimento di Matematica Pura ed Applicazioni,\\
Universit\`{a} di Milano Bicocca, 20125 Milano, Italy\\
{giovanni.ortenzi@unimib.it} }
\maketitle
\abstract{Changes of type transitions for the two-component hydrodynamic type systems are  discussed.  It is shown that 
these systems generically assume the Jordan form (with $2 \times 2$ Jordan block) on the transition line with hodograph equations
becoming parabolic. Conditions which allow or forbid 
the transition from hyperbolic domain to elliptic one are discussed. Hamiltonian systems and their special subclasses and equations, like 
dispersionless nonlinear Schr\"odinger, dispersionless Boussinesq, one-dimensional isentropic  gas dynamics equations and nonlinear 
wave equations are studied. Numerical results concerning the crossing 
of transition line for the dispersionless Boussinesq equation are presented too.}
\section{Introduction}
Differential equations and systems of mixed type always have attracted a great interest due to the presence of both hyperbolic and elliptic regimes, 
possibility of transition to each other and numerous interpretations of such transitions in various fields of mathematics, 
physics
and applied science (see e.g. \cite{CH}-\cite{Otw}). \par
Study of the properties of the systems of quasi-linear equations near the transition line (referred also as sonic line, parabolic line 
or hyperbolic-elliptic boundary) is of particular interest due to the connection with the problem of nonlinear stability of systems of mixed type.
The results  obtained  in the recent papers \cite{M5,T6,CT} contributed significantly to understanding and clarifying the situation. \par
 On the other hand some assumptions made, for instance, in \cite{CT} seems to be rather restrictive. In particular,
 the calculations made in \cite{CT} are based on the hypothesis that the matrix $V$ for the 
 hydrodynamic type systems 
 \begin{equation}
 \label{gensysvec}
  \vec{u}_t= V(\vec{u}) \vec{u}_x
 \end{equation}
``is degenerated yet diagonalizable'' on sonic line \cite{CT}. \par
There are, however, number of systems for which it is not the case.
The simplest example is provided by the well-known one-layer Benney system (or dipersionless 
nonlinear Schr\"odinger equation (dNLS))
\begin{equation}
 \begin{split}
  &u_t=uu_x+v_x\,  ,\\
  &v_t=vu_x+uv_x.
 \end{split}
 \label{dNLS}
\end{equation}
Characteristic speeds for (\ref{dNLS}) are $\lambda_\pm=u \pm \sqrt{v}$ and the transition line is given by equation $v(x,t)=0$. On the 
transition line the matrix $V$ 
takes the form 
\begin{equation}
\label{gascompr}
V_0=\left(
 \begin{array}{cc}
  u &1 \\ 0 & u
 \end{array}
\right)
 \end{equation}
which is obviously non-diagonalizable. Another example is provided by the dispersionless Boussinesq (dB) equation
\begin{equation}
\label{dBi}
 u_{tt}=\frac{1}{2}(u^2)_{xx} 
\end{equation}
or the system
\begin{equation}
\label{dB2i}
\begin{split}
 & u_t=v_x \, ,\\
 & v_t=uu_x\, .
\end{split}
\end{equation}
In this case the matrix $V$ on the transition line is
\begin{equation}
\label{dB2iJ}
V_0=\left(
 \begin{array}{cc}
  0 & 1 \\ 0 & 0
 \end{array}
\right) \, , 
\end{equation}
i.e. the Jordan block with zero eigenvalue. \par
The appearance of Jordan blocks in the these examples is a clear manifestation  of the generic structure
for the hydrodynamics type systems on the transition line. \par
These two systems represent two different classes  of hydrodynamic systems of mixed type. For dNLS equation (\ref{dNLS})
hyperbolic domain ($v>0$) is  separated from the elliptic one ($v<0$). For the  dB equation (\ref{dBi})
the transition line can be crossed and solutions from the hyperbolic domain
($u>0$) can pass to the elliptic domain ($u<0$). \par 
In the present paper these phenomenons are studied for the two-component
systems (\ref{gensysvec}) of mixed type. 
 It is shown  that generically a two-component system (\ref{gensysvec}) at the transition line
is of Jordan form. 
Hodograph equations are manifestly parabolic on the 
transition  line. This parabolic regime separates the hyperbolic domain describing wave propagation  and elliptic
domain containing quasi-conformal mapping. Conditions  under which solutions of the system (\ref{gensys})  may belong to both  hyperbolic
and elliptic domain or avoid the crossing of the transition line are discussed. \par
Hamiltonian systems are considered in detail as illustrative examples.  It is shown that the presence of the
Jordan block on the transition line is a typical behavior of Hamiltonian systems. \par
It is also  that in the generic case the characteristics in $(u,v)$ plane (simple waves) have universal behavior
\begin{equation}
 v-v_0 \sim (u-u_0)^{3/2}
\end{equation}
near the point $(u_0,v_0)$ of contact with the transition line. The dB equation  is characteristic representative of  such a behavior.
Particular classes of Hamiltonian systems, including gasdynamics equations and nonlinear wave equations are considered.\par  
Numerical results for the dB equation showing 
the particularities of crossing of the transition line are presented too. \par
The paper is organized as follows. Some basic well-known results for the $2 \times 2$  system (\ref{gensys}), including hodograph
equations are given in  Section {\ref{sect-gen}}. Behavior of the system  (\ref{gensys})  on the transition line,
its Jordan form, and parabolic character  are considered in Section \ref{sect-TL}. In section \ref{sect-ell} it is shown the behavior of
the system near the transition line from the elliptic side.
Necessary and sufficient conditions which allows 
or forbid  the crossing of the transition line are discussed in Section \ref{sect-cross}. These results applied to general Hamiltonian systems are 
presented  in Section \ref{sect-Ham}. Special classes of Hamiltonian systems and, in particular, equations of motion for
isentropic gas equations and nonlinear wave equations are considered in Sections \ref{sect-exe} and   and \ref{sect-NLW}. Some numerical results for 
the dB equation near to the transition line are presented in Section \ref{sect-num}. 
\section{General formulae}
\label{sect-gen}
We will consider two component quasi-linear system of mixed type of first order
\begin{equation}
\label{gensys}
 \binom{u_t}{v_t}= \left(
 \begin{array}{cc}
  A &B \\ C & D
 \end{array}
\right)  \binom{u_x}{v_x}
\end{equation}
where $A,B,C,D$ are certain real functions of $u$ and $v$ and subscript denotes derivatives. For convenience we will recall here some basic 
known facts (see e.g. \cite{CH,Whi,L-VI}). Generically the matrix
\begin{equation}
\label{mat-v}
 V=  \left(
 \begin{array}{cc}
  A &B \\ C & D
 \end{array}
\right)
\end{equation}
has two distinct eigenvalues given by
\begin{equation}
 \lambda_\pm= \frac{A+D \pm \sqrt{(A-D)^2+4BC} }{2}\, .
\end{equation}
If $\Omega \equiv (A-D)^2+4BC > 0 $ the system is hyperbolic, while at $\Omega<0$ it is elliptic. 
In this paper we will assume that $\Omega(u,v)$ is a smooth function of $u$ and $v$. So, the hyperbolic 
and elliptic domains are separated by the transition line give by the equation
\begin{equation}
\label{HEline}
 \Omega(u(x,t),v(x,t))=0 \, .
\end{equation}
Classical hodograph equations for the system (\ref{gensys}) is
\begin{equation}
\label{genhodo}
 \binom{x_u}{x_v}= \left(
 \begin{array}{cc}
  -D & C \\ B & -A
 \end{array}
\right)  \binom{t_u}{t_v}.
\end{equation}
As a consequence, the variables $t$ and $x$ obey the second order equations
\begin{equation}
\label{teqn}
 C t_{vv}+2\ \frac{A-D}{2}t_{uv} -B t_{uu} -(B_u+D_v)t_u +(A_u+C_v)t_v=0
\end{equation}
and
\begin{equation}
\label{xeqn}
 \left(\frac{Ax_u+Cx_v}{AD-BC} \right)_v-\left( \frac{B x_u+D x_v}{AD-BC} \right)_u=0\, .
\end{equation}
If the system (\ref{gensys}) has a conservation equation $Q_t=P_x$ then $Q$ obeys the equation
\begin{equation}
\label{Qeqn}
 C Q_{vv}+ ({A-D})Q_{uv} -B Q_{uu} -(B_u-A_v)Q_u +(-D_u+C_v)Q_v=0\, . 
\end{equation}
In the hyperbolic domain there are two real Riemann invariants $r_+$ and $r_-$ such that the system (\ref{gensys}) is equivalent to
\begin{equation}
\label{Rinvform}
 {r_\pm}_t= \lambda_\pm {r_\pm}_x
\end{equation}
with two distinct characteristic speeds $\lambda_+$ and $\lambda_-$. In the elliptic domain $\lambda_+$ and $\lambda_-$ are complex-conjugate to 
each-other and one has the single complex equation
\begin{equation}
 {r_+}_t= \lambda_+ {r_+}_x
\end{equation}
with $r_-$ being the complex conjugate to $r_+$. Riemann invariants obey to the system
\begin{equation}
\begin{split}
& (A-\lambda_\pm) {r_\pm}_u+C {r_\pm}_v=0, \\
& B {r_\pm}_u+(D-\lambda_\pm) {r_\pm}_v=0 \, .
\end{split}
\end{equation}
Only two among these equations are independent say
\begin{equation}
  B\, {r_\pm}_u+(D-\lambda_\pm)\, {r_\pm}_v=0
\end{equation}
or
\begin{equation}
\label{ruv}
   {r_\pm}_u=\frac{A-D\pm\sqrt{\Omega}}{2B}\, {r_\pm}_v \, .
\end{equation}
Equations (\ref{ruv}) are diagonal form of the hodograph equations (\ref{genhodo}) rewritten as
 \begin{equation}
\label{genhodo-tx}
 \binom{t_u}{x_u}= \frac{1}{B}\left(
 \begin{array}{cc}
  A & 1 \\ BC-AD & -D
 \end{array}
\right)  \binom{t_v }{x_v}.
\end{equation}
Indeed the eigenvalues $\mu_\pm$ of the matrix present in (\ref{genhodo-tx})
are
\begin{equation}
 \mu_\pm= \frac{A-D\pm\sqrt{\Omega}}{2B}\, .
\end{equation}
The characteristics for equation (\ref{ruv}) are defined by the equation
\begin{equation}
\label{SW}
\left(\frac{\D v}{ \D u}\right)_\pm =-\mu_\pm \, .
\end{equation}
Riemann invariants are constants along these characteristics $v_\pm$ in the hodograph space. \par
Equations (\ref{SW}) are those which define simple waves for the system (\ref{gensys}). The simple waves 
are the hodograph counterpart of the usual characteristics $\left(\frac{\D x}{\D t}\right)_\pm=-\lambda_\pm$ in the space $(x,t)$.
 We note also that the components $y_1$ and $y_2$ of the eigenvector {\textit{\textbf y}} of the matrix V in (\ref{mat-v}) 
 on the transition line
 obey the equation
\begin{equation}
\label{eigen-v}
(A-D)y_1+2By_2=0\, . 
\end{equation}
Hydrodynamic type systems  and, in particular, the system (\ref{gensys}) exhibit one more important phenomenon, the so-called gradient catastrophe,
i.e. unboundedness of derivatives of $u$ and $v$ at finite $x$ and $t$ while $u(x,t)$ and $v(x,t)$ remain bounded (see e.g. \cite{Whi}). Interference of the 
gradient catastrophe and crossing the transition line is a rather complicated problem. To simplify the analysis we will assume in the rest of 
the paper 
that the solutions of the system (\ref{gensys}) avoid gradient catastrophe, at least, before the crossing of the transition line (if so).
\section{Transition line and Jordan form}
\label{sect-TL}
Study of the behavior of systems of mixed type on the transition line and nearby is fundamental for understanding their properties. 
The system (\ref{gensys}) can be  of mixed type only when  $B C < 0$. In the case $BC\geq0$  (including symmetric matrices) it is hyperbolic except 
the degeneration at the set of points defined by the equation $A=D$ and $B=0$ (or $C=0$) in generic case or at the line in the case $A=D$ 
and $B=C=0$. \par

It was already shown in the Introduction that on the transition lines the dNLS and dB equations assume the special form with Jordan blocks. 
These results   can be obtained as the limit, performed accurately, of 
the equations (\ref{Rinvform}) for Riemann invariants 
when a solution $(u,v)$ approaches the transition line. 
In the dNLS equation (\ref{dNLS}) case 
\begin{equation}
 \lambda_{\pm}=u \pm \sqrt{v}, \qquad  r_{\pm}=u \pm 2\sqrt{v}
\end{equation}
and the transition line is given by the equation $v(x,t)=0$. Equations for the Riemann invariants in terms of $u$ and $v$ are
\begin{equation}
 (u \pm 2\sqrt{v})_t=(u \pm \sqrt{v})(u \pm 2\sqrt{v})_x
\end{equation}
or
\begin{equation}
 u_t \pm \frac{1}{\sqrt{v}}v_t=uu_x \pm \sqrt{v} u_x \pm \frac{1}{\sqrt{v}} u v_x + v_x\, . 
 \end{equation}
In the limit $v \to 0$ the two leading order terms $v^{-1/2}$ and $v^0$ give the system 
\begin{equation}
 u_t = uu_x +v_x\, , \qquad v_t = u v_x  \, .
\end{equation}
This system has matrix $V$ given by (\ref{gascompr}), that is the Jordan form 
with $\lambda = u$. \par
In the dB equation (\ref{dB2i}) case
\begin{equation}
 \lambda_\pm=\pm u^{1/2}\, , \qquad r_\pm=v\pm \frac{2}{3}u^{3/2}\, ,
\end{equation}
and the transition line is given by the equation $u(x,t)=0$.
So equations (\ref{Rinvform}) in terms of $u$ and $v$ are
\begin{equation}
 (v\pm \frac{2}{3}u^{3/2})_t=\pm u^{1/2}(v\pm \frac{2}{3}u^{3/2})_x
\end{equation}
or
\begin{equation}
 v_t\pm u^{1/2}u_t=\pm u^{1/2} v_x + u u_x\, .
\end{equation}
When $x$ and $t$ approaches the transition line $u(x,t)=0$ in the leading orders $u^0$ and $u^{1/2}$ one gets 
\begin{equation}
 v_t=0\, , \qquad  u_t=v_x\, ,
\end{equation}
i.e. the Jordan form with $\lambda=0$. \par
In the general case the matrix $V$ for the system (\ref{gensys}) apparently is not of the Jordan block form on the transition line 
$\Omega$ (\ref{HEline}).  It   can be parameterized at $C\neq 0$ as 
\begin{equation}
\label{mat-v0}
 V_0\equiv V|_{\Omega=0}= \left(
 \begin{array}{cc}
  \lambda + \sqrt{-BC} & B \\ C & \lambda - \sqrt{-BC}
 \end{array}
\right) 
\end{equation}
where $\lambda=\lambda_+=\lambda_-=(A+D)/2$.
Such a matrix has the form
\begin{equation}
 V_0= \lambda  \left(
 \begin{array}{cc}
 1 & 0\\ 0 & 1
 \end{array}
\right) +N
\end{equation}
where $N$ is the general $2\times 2$ nilpotent matrix. \par
It is straightforward to check that there exists a two parameter family of invertible 
matrices $P$ such that
\begin{equation}
\label{transfV}
 P V_0 P^{-1}= \left(
 \begin{array}{cc}
 \lambda & 1\\ 0 & \lambda
 \end{array}
\right).
\end{equation}
The family $P$ at $C\neq 0$ is given by
\begin{equation}
\label{transmatP}
P=a
\left(
\begin{array}{cc}
 -b  & 1+\frac{\sqrt{-B C}}{C} b  \\ & \\
 C & -\sqrt{-B C} 
\end{array}
\right)\, ,
\end{equation}
where $a=a(u,v)$ and $b=b(u,v)$ are two arbitrary functions. 
In the case of $C=0$ and $B \neq 0$ the matrix $P$ becomes
\begin{equation}
\label{transmatP-deg}
 P=a
\left(
\begin{array}{cc}
 1  &  b  \\
 0 & B 
\end{array}
\right)\, ,
\end{equation}
where $a=a(u,v)$ and $b=b(u,v)$ are still two arbitrary functions.
Thus is all non diagonal cases the matrix $V_0$ is equivalent
to a Jordan block, 
and the system (\ref{gensys}) is equivalent to
\begin{equation}
 \label{ontl}
 P \binom{u_t}{v_t} =  \left(
\begin{array}{cc}
 \lambda  &  1  \\
 0 & \lambda \\
\end{array}
\right)  P \binom{u_x}{v_x}\, . 
\end{equation} \par
 In our construction the systems in the Jordan forms or the system (\ref{ontl}) arise on the transition line only.
 In the paper \cite{kodakono} the system (\ref{gensys}) with matrix $V$ given by (\ref{gascompr}) on the whole plane $(x,t)$ and its 
 multi-component analogs with Jordan blocks has been derived via the confluence process for the Lauricella-type functions 
 associated with Grassmannians Gr$(2,5)$ and Gr$(2,n)$.\par
 Let us consider the system (\ref{gensys}) with the matrix $V=V_0$ given by (\ref{mat-v0}). It is parabolic on the plane (x,t). In this case 
 there are variables $u^*, v^*$  such the system (\ref{ontl})  takes the form  
\begin{equation}
\label{Jordform}
 \binom{u^*_t}{v^*_t}= \left(
 \begin{array}{cc}
  \lambda & 1 \\ 0 & \lambda
 \end{array}
\right)  \binom{u^*_x}{v^*_x}\, . 
\end{equation}
which will be referred as the Jordan form. The relation between the Jordan variables $u^*,v^*$ and the original ones $u,v$ can be found solving 
the equations
\begin{equation}
 \binom{u^*_t}{v^*_t}=P\binom{u_t}{v_t}, \qquad \binom{u^*_x}{v^*_x}=P \binom{u_x}{v_x} \, .
\end{equation}
These equations can be solved only if one finds the suitable integrating factors $a$ and $b$ which must be chosen such that
\begin{equation}
\D u^* =  a b \,  \D u+  a \left(1+\frac{\sqrt{-B C}}{C} b \right) \,  \D v \, , \qquad
\D v^* = - a  \,  \D u- a \sqrt{-B C}  \,   \D v  \, ,
\end{equation}
which fix $a$ and $b$ thanks to the compatibility conditions
\begin{equation}
 (a b)_v=  \left(a +\frac{a \sqrt{-B C}}{C} b \right)_u, \qquad  a_v  =   (a \sqrt{-B C})_u \, .
\end{equation}
Therefore the Jordan variables in parabolic systems
play a role similar to the Riemann invariants for the standard diagonalizable case.
\section{Elliptic domain}
\label{sect-ell}
At the hyperbolic domain solution of the system (\ref{gensys}) describe wave motions. Properties of solutions of the system (\ref{gensys}) 
in the elliptic domain are quite different. Their treatment as the function defining quasi-conformal mappings is one of the possible 
interpretations
\cite{Lav}. Indeed equation (\ref{Rinvform}) can be rewritten as \cite{KOsapm} 
\begin{equation}
 \label{Rinvform-ell}
 r_{\overline{z}}= \frac{1+i\lambda}{1-i\lambda} r_z
\end{equation}
where $\lambda = \lambda_+ = (A+D + \sqrt{\Omega})/2$, $r=r_+$,  $z=x+it$ and the overline stands for complex conjugate. 
Such equations are known as Beltrami nonlinear equations 
\cite{Alf,Boj}. 
A solution of (\ref{Rinvform-ell}) defines a quasi-conformal
mapping $r:\, (z,\overline{z}) \to (r,\overline{r})  $ if the complex dilation $\mu=\frac{1+i\lambda}{1-i\lambda}$ obeys the condition
\begin{equation}
\label{conf-cond}
 |\mu| = \left\vert \frac{1+i\lambda}{1-i\lambda} \right\vert < 1 \, .
\end{equation}
Using the explicit form of $\lambda$, in the region $\Omega < 0$ we have
\begin{equation}
 \mu= \frac{2+  i(A+D) -  \sqrt{- \Omega} }{2-  i(A+D) +  \sqrt{- \Omega}}\, .
\end{equation}
It is easy to see that condition (\ref{conf-cond}) is always satisfied when $\Omega<0$. So any solution  of the system
(\ref{gensys}) in the elliptic domain  defines quasi-conformal mapping. \par
At the transition line $\Omega = 0$ 
\begin{equation}
 |\mu|= \left\vert \frac{2+  i(A+D)  }{2-  i(A+D) } \right\vert =1 
\end{equation}
and the quasi conformal mappings degenerates. For instance it maps the unit circle in the plane $(z, \overline{z})$
 in the degenerate ellipsis in the plane $(r,\overline{r})$ with ratio of major and minor axes going to infinity
(see e.g. \cite{Alf}). \par
In the hodograph space, equations (\ref{ruv}) are equivalent to the linear Beltrami equation
\begin{equation}
\label{confmapuv}
 r_{\overline{w}}=\frac{1+i\tilde{\lambda}(w,\overline{w})}{1-i\tilde{\lambda}(w,\overline{w})} r_w
\end{equation}
where $w=v+iu$ and $\tilde{\lambda}=\frac{A-D+i\sqrt{-\Omega}}{2B}$. Similar to the calculations presented before, one shows that the condition
(\ref{conf-cond}) is always satisfied in the elliptic domain and so any solution of the equation (\ref{confmapuv}) defines
a quasi-conformal mapping $(w,\overline{w})\to (r,\overline{r})$. On the transition line $\Omega=0$ again 
$\left| \frac{1+i\tilde{\lambda}}{1-i\tilde{\lambda}} \right|=1$ and  quasi-conformal mappings become  
singular. \par 
So, both in the elliptic and hyperbolic domains  solutions of the mixed system (\ref{gensys}) exhibit particular behavior
when they approach the transition line $\Omega=0$. Namely, approaching the transition line from the hyperbolic side waves become unstable
(see e.g. \cite{Cra}) converting into mess, governed by parabolic equations  and transforming into quasi-conformal
mappings dynamics beyond the transition line. Approaching the transition line from the elliptic side, the quasi-conformal mappings
degenerate into singular ones  with $|\mu|=1$ which maps the two dimensional domains in $\mathbb{C}$ into quasi one-dimensional
ones. Beyond the transition line these quasi one-dimensional objects are transformed into moving waves.
\section{Transition line and its crossing}
\label{sect-cross}
Since Riemann invariants are constants along characteristics (real or complex) the problem whether or not a transition  line
can be crossed is reduced to the study of respective properties of characteristics and transition lines (see e.g. \cite{M5,T6,CT}). Comparison
of the formulae for characteristics and transition line in the original variables $(t,x)$ and hodograph variables $(u,v)$ (see e.g.
formulae (\ref{HEline}) and (\ref{SW}) clearly indicates that the latter ones is more appropriate for our purpose. The use of simple
waves in  \cite{T6} provides us another support of such observation. \par
In the hodograph space $(u,v)$ the characteristics and transition lines are given by formula (\ref{SW}) and (\ref{HEline}) respectively.
Let us begins with hyperbolic domain and let us assume that the derivatives involved are bounded. Thus we have two families of  plane 
characteristic lines (ChL) in the hodograph space ($B \neq 0$)
\begin{equation}
\label{chl}
 \mathrm{ChL}: \qquad \left(\frac{\D v}{ \D u}\right)_\pm =-\frac{A-D\pm \sqrt{\Omega}}{2B}\, ,
\end{equation}
and a single transition line (TL)
\begin{equation}
 \label{tl}
\mathrm{TL}: \qquad  \Omega(u,v)=0\, .
\end{equation}
The two simplest cases are: 1) the two families (\ref{chl}) do not have common points with (\ref{tl}) and 2) 
they coincide at least on some interval. In the latter case, on the transition line  one has the equations 
\begin{equation}
 \frac{\D v}{\D u}= \frac{D-A}{2B}, \qquad \Omega=0\, ,
\end{equation}
which should be equivalent to each other. 
Since on the transition line 
\begin{equation}
 \D \Omega = \Omega_u \D u +\Omega_v \D v =0\, ,
\end{equation}
the necessary condition for this is given by (if $\Omega_v \neq 0$)
\begin{equation}
\label{crosscond}
 \frac{D-A}{2B}+  \frac{ \Omega_u}{ \Omega_v}=0\, .
\end{equation}
Obviously in both cases the transition  from the hyperbolic domain to the elliptic one is impossible. \par
Another simple case corresponds to the 
transversal intersections of ChL and TL. To derive the corresponding condition it is sufficient to consider these lines at points of 
intersection. Two characteristic touch each other and at the point on TL their tangents are (assuming that both curves are smooth)
\begin{equation}
 \frac{\D v}{\D u}\Big\vert_\mathrm{ChL}= \frac{D-A}{2B}\, .
\end{equation}
Tangent to the TL at the same point is given by (at $\Omega_v \neq 0$) 
\begin{equation}
 \frac{\D v}{\D u}\Big\vert_\mathrm{TL} = - \frac{ \Omega_u}{ \Omega_v}\Big\vert_\mathrm{TL}\, .
\end{equation}
Characteristic and transition line cross transversally (with angle $\neq 0$) if 
\begin{equation}
 \frac{\D v}{\D u} \Big\vert_{\mathrm{ChL}} \neq \frac{\D v}{\D u}\Big\vert_{\mathrm{TL}}\, ,
\end{equation} 
i.e.
\begin{equation}
 \label{transvcond}
\left( \frac{D-A}{2B}+ \frac{ \Omega_u}{ \Omega_v}\right)_\mathrm{TL} \neq 0\, .
\end{equation}
Thus, if condition (\ref{transvcond}) is satisfied, the transition from the hyperbolic domain to the elliptic one is not forbidden. \par
\begin{figure}
\centering
{\includegraphics[height=5cm]{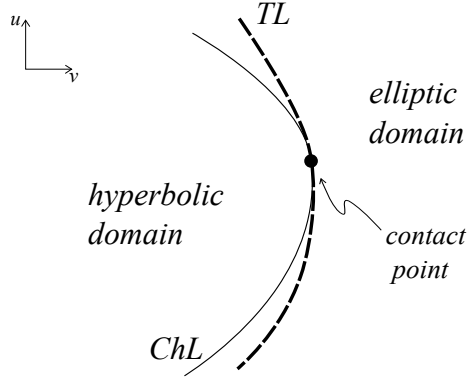}}
\caption{A configuration of characteristic (solid) and transition line (dashed) in the $(u,v)$-space 
where the hyperbolic-elliptic transition in the contact point is forbidden.}
\label{fig-notrans}
\end{figure}

There are eight other possibilities. First four are given by the figure \ref{fig-notrans} and its reflections each of two curves with respect 
the straight line of common tangent at the point $(u_0,v_0)$.
In these four cases characteristic lines touches the transition line at the point $(u_0,v_0)$ and then turns back to the  hyperbolic domain.
So the transition is forbidden. At the point $(u_0,v_0)$ tangents of both side coincides and so
\begin{equation}
\label{nonccond}
\left(  \frac{\D v}{\D u} \right)_{\mathrm{ChL}} \Big\vert_{(u_0,v_0)}-
\left(  \frac{\D v}{\D u} \right)_{\mathrm{TL}} \Big\vert_{(u_0,v_0)}=
\left( \frac{D-A}{2B}+ \frac{ \Omega_u}{ \Omega_v}\right)_{(u_0,v_0)} =0\, .
\end{equation}
The fact of non-crossing is invariant under the transformation of coordinates. Choosing the coordinates $(u,v)$ near to the point 
$(u_0,v_0)$ in such a way that the axes $v=0$ coincides with the common tangent, it is not difficult to show in all four cases that 
the difference
\begin{equation}
\label{DT}
 \Delta T \equiv \left(  \frac{\D v}{\D u} \right)_{\mathrm{ChL}}-\left(  \frac{\D v}{\D u} \right)_{\mathrm{TL}}
\end{equation}
changes sign passing the point $(u_0,v_0)$. \par
\begin{figure}
\centering
{\includegraphics[height=5cm]{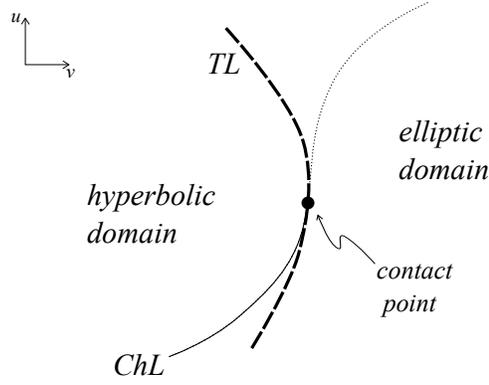}}
\caption{A configuration of characteristic (solid) and transition line (dashed) in the $(u,v)$-space 
where the hyperbolic-elliptic transition in the contact point requires 
a deeper analysis and it could not be forbidden. In the elliptic region the characteristic line (here depicted as dotted) becomes complex.}
\label{fig-trans}
\end{figure}

Second four cases are given by figure (\ref{fig-trans}) and three others possible are obtained by reflection each of two curves with respect the 
straight line of common tangent at the point $(u_0,v_0)$.
In these cases characteristic touches the transition line at the point $(u_0,v_0)$ and then pass into elliptic domain with characteristic speeds becoming complex.
Thus, the transition is not forbidden. \par
Thus in the eight cases considered above the behavior of $\Delta T$ near the point of touch $(u_0,v_0)$ distinguishes the cases of crossing 
and non-crossing. So we conclude that the transition from the hyperbolic to the elliptic domain is not possible if either $\Delta T|_{\Omega=0}=0$
on some interval of the transition line of $\Delta T|_{\Omega(u_0,v_0)=0}=0$ at some point on TL and $\Delta T$ changes sign passing from one 
side of the touch point $(u_0,v_0)$ to another. In particular, the comparison of the condition (\ref{nonccond}) rewritten as
\begin{equation}
\label{nonccond-v2}
\left( (D-A)\Omega_v+2B \Omega_u \right)|_{(u_0,v_0)}=0\, 
\end{equation}
and the relation (\ref{eigen-v}) shows that the eigenvector of the matrix $V$
 corresponding to the double eigenvalue $\lambda$ is tangent to the transition line at the point $(u_0,v_0)$ in agreement with necessary
 condition of non-crossing  (nonlinear stability) proposed in \cite{M5}.  On the other hand if $\Delta T|_{\Omega=0} \neq 0$ or
 $\Delta T|_{\Omega(u_0,v_0)=0} = 0$ and $\Delta T$ does not change sign at the touching point $(u_0,v_0)$ the conditions of non-crossing 
 are not satisfied and the transition from the hyperbolic domain to the elliptic one is not forbidden. \par
 In the analysis presented above it was assumed that all derivatives including $\frac{\D^2 v}{\D u^2}$ are bounded. The cases of possible 
 unboundedness require special consideration. To clarify the point let us consider the system 
 \begin{equation}
  \label{strange-NLW}
  \binom{u_t}{v_t}= \left(\begin{array}{cc}
                            0&1\\f(v)&0
                          \end{array} \right)    \binom{u_x}{v_x}
 \end{equation}
where $f(v)=v^\alpha$ and $\alpha=2n+1$ or $\alpha=1/(2n+1)$ with $n=0,1,2,3,\dots$ . In this case the transition line is given by $v=0$ 
and the characteristics in the hodograph space are defined by the equation 
\begin{equation}
 \frac{\D v}{\D u}= \mp v^{\alpha/2}\, .
\end{equation}
So, the characteristics are given by the  lines 
\begin{equation}
\label{SW-strange}
 u \pm \frac{2}{2-\alpha} v^{(2-\alpha)/2}=u_0=const\, , \qquad v \neq 0
\end{equation}
with arbitrary $u_0$ and by the straight line $v=0$. The transition line $v=0$  is then (degenerate)  characteristics. The behavior 
of other characteristics is quite different for $\alpha>2$ and $0<\alpha<2$. In the case $\alpha=2n+1$ and  $n \geq 1$, $v(u)$ has a singularity
at $u=u_0$ and it it may touch the transition line only at the infinity $u \to \infty$. So the transition from hyperbolic to elliptic domain 
is not possible. For $\alpha =\frac{1}{2n+1}$, the characteristics (\ref{SW-strange}) touch the transition line at finite point
$(u_0,0)$ where $\frac{\D v}{ \D u} \Big\vert_{u_0,0}=0$. Since
\begin{equation}
 \frac{\D^2 v }{\D u^2}= \left(\frac{2-\alpha}{2}(u-u_0)\right)^\frac{2(\alpha-1)}{2-\alpha}=\frac{\alpha}{2}v^{\alpha-1}
\end{equation}
the characteristics have completely different behavior in the cases $\alpha=1$ and $\alpha<1$. For $\alpha=1$ one has 
\begin{equation}
  \frac{\D^2 v }{\D u^2} \Big\vert_{(u_0,0), \alpha=1}= const\, . 
\end{equation}
So characteristics approach smoothly the transition line and, hence, the transition is not possible. Note that at $\alpha=1$ the system 
(\ref{strange-NLW}) represent the equation 
\begin{equation}
 (\log v)_{tt}=v_{xx}\, .
\end{equation}
In contrast, for $\alpha<1$ ($n=1,2,3,\dots$) the normal to the characteristic, i.e. velocity $\frac{\D^2 v }{\D u^2}$ with which
characteristic approaches (in normal direction) the transition line $v=0$ grows to infinity (at the point $(u_0,0)$). Such a behavior
allows us to suggest that the characteristics may jump across the line $v=0$ and transition from the hyperbolic domain to the elliptic
one $(v<0)$ could be possible. The numerical results of the system (\ref{strange-NLW}) with $f=v^{1/3}$ ($\alpha=1/3$) presented in the paper 
\cite{M5} support this observation. Another indication that the system (\ref{strange-NLW}) with $f=v^\alpha$, $\alpha<1$ has
has rather special properties is provided by equations (\ref{teqn}) and (\ref{xeqn}), i.e.
\begin{equation}
 v^\alpha t_{uv}-t_{uu}+a v^{\alpha-1}t_v=0
 \end{equation}
and similar equation for $x$. At the transition line $v=0$ they are singular equations of parabolic type. \par
Analysis of the systems which have properties similar to those of system (\ref{strange-NLW}) with $f=v^\alpha$, $\alpha<1$
requires a separate study which will be performed elsewhere. Possibility of transition from elliptic to hyperbolic domain, corresponding 
conditions and associated quasi-conformal mapping are of interest too.
 These problems will be considered in a separate publication.
\section{Hamiltonian systems}
\label{sect-Ham}
In this and next sections we will consider some concrete classes of equations (\ref{gensys}). 
We begin with the Hamiltonian systems which for 
two component case can always be locally put in the form \cite{DN,Dub}
\begin{equation}
 \binom{u_t}{v_t} = \left( \begin{array}{cc} 0 &\partial_x \\ \partial_x &0 \end{array} \right) \binom{h_u}{h_v}
\end{equation}
with Hamiltonian $H=\int h(u,v) \, \D x$. So, the system (\ref{gensys}) is
\begin{equation}
\label{gensys-Ham}
 \binom{u_t}{v_t} = \left( \begin{array}{cc} h_{uv} &h_{vv} \\ h_{uu} &h_{uv} \end{array} \right) \binom{u_x}{v_x}\, .
\end{equation}
In this case $\Omega=4 h_{uu}h_{vv}$. Equation (\ref{teqn}) is
\begin{equation}
\label{teqn-Ham}
 h_{uu} t_{vv}- h_{vv} t_{uu}-2 h_{uvv} t_{u}+2 h_{uuv} t_{v}=0\, ,
\end{equation}
while characteristics in the hodograph space (simple waves) are defined by the equation
\begin{equation}
\label{SW-Ham}
 \left( \frac{\D v}{\D u}\right)_\pm=\mp \sqrt{\frac{h_{uu}}{h_{vv}}}\, .
\end{equation}
Transition line is given by 
\begin{equation}
 \label{TL-Ham}
 h_{uu}h_{vv}=0\, ,
\end{equation}
assuming that $h_{uu}h_{vv}$ may change sign. \par
In order to deal with generic non-diagonalizable case we defined the transition line as
\begin{equation}
\label{uu-branch}
 h_{uu}=0, \qquad h_{vv}|_{h_{uu}=0}\neq 0\, .
\end{equation}
On the transition line the matrix $V$ is equivalent to the Jordan block
$\left( \begin{array}{cc} \lambda & 1 \\ 0 & \lambda \end{array} \right)$ \, 
with  $\lambda=h_{uv}|_{h_{uu}=0}$. The transformation matrix $P$ is given by (\ref{transmatP-deg}) 
with $B= h_{vv}|_{h_{uu}=0}$.
We note that at  points where $h_{vv}=0$ and ${h_{uu}=0}$ on the transition line  the matrix  $V$ is degenerated to a constant diagonal matrix. 
\par
When it holds (\ref{uu-branch}) one also has
\begin{equation}
 \Delta T= \frac{h_{uuu}}{h_{uuv}}\Big\vert_{h_{uu}=0}\, .
\end{equation}
Thus in generic case with $h_{uuu}|_{h_{uu}=0} \neq 0$ we have
\begin{equation}
 \Delta T \neq 0 
\end{equation}
and transition from the hyperbolic domain $h_{uu}h_{vv}>0$ to the elliptic one is not forbidden. \par
Both dNLS and dB equations are Hamiltonian ones (see e.g. \cite{Whi,DN,Dub} ) with respectively
\begin{equation}
\label{Ham-dNLS}
 h_{\mathrm{dNLS}} = \frac{1}{2}v^2 +\frac{1}{2}vu^2\, , 
\end{equation}
and
\begin{equation}
\label{Ham-dB}
h_{\mathrm{dB}} = \frac{1}{2}v^2 +\frac{1}{6}u^3\, . 
\end{equation}
Now let us study the behavior of characteristics in $(u,v)$ plane (simple waves) near the point $(u_0,v_0)$ of contact with the transition line 
$h_{uu}=0$ and $h_{vv}\neq 0$ for general Hamiltonian systems (\ref{gensys-Ham}).\par
Expanding the right hand side of (\ref{SW-Ham}) near the point $(u_0,v_0)$ and assuming that $h_{uuu} \neq 0$ and the derivatives envolved 
are finite, one obtains
\begin{equation}
\label{ChLnearTL}
 \left(\frac{\D v}{\D u}\right)_\pm=\pm \sqrt{a(u-u_0)+b(v-v_0)}, \qquad \mathrm{with} \quad a=\frac{h^0_{uuu}}{h^0_{vv}} \neq 0
 \quad b=\frac{h^0_{uuv}}{h^0_{vv}},
\end{equation}
where $f^0\equiv f(u_0,v_0)$. For infinitesimal $\delta u= u-u_0$ and $\delta v= v-v_0$   equation (\ref{ChLnearTL}) takes the form
 \begin{equation}
 \label{ChLnearTL-loc}
  \left(\frac{\delta v}{\delta u}\right)^2=a \delta u+b \delta v\, .
 \end{equation}
Hence
\begin{equation}
 \delta v= \frac{1}{2} b (\delta u)^2 \pm \sqrt{\frac{b^4}{2} (\delta u)^4 + a (\delta u)^3}\, .
\end{equation}
So at $\delta u\to 0$, $\delta v \to 0$ one has at the leading order
\begin{equation}
\label{ChLnearTL-gen}
 v-v_0 \simeq \pm \sqrt{\frac{h^0_{uuu}}{h^0_{vv}}} (u-u_0)^{3/2}\, .
\end{equation}
This formula gives us the universal behavior of $(u,v)$ characteristics near the transition line for general Hamiltonian system (\ref{gensys-Ham})
in the generic case $h^0_{uuu} \neq 0$. The simplest and characteristic example of such a behavior is provided by the dB equation 
for which $h^0_{uuu}=1$. It should be also noted that the fact, that for the general stationary plane motion of compressible gas 
(described by Chaplygin equation) the behavior of characteristics near sonic line is given by the formula (\ref{ChLnearTL-gen}), has been known 
for a long time (see e.g. \cite{L-VI}, \S 118).\par
In particular cases the behavior of characteristics near to the transition line is quite different. If at the transition point $(u_0,v_0)$
also $h^0_{uuu}=0$, then, instead of equation (\ref{ChLnearTL-loc}) one has
\begin{equation}
 \label{ChnearTl-loc-nongen}
 \left(\frac{\delta v}{\delta u}\right)^2= b \delta v +c(\delta u)^2 +d \delta u\, \delta v +f (\delta v)^2
\end{equation}
where $c,d$ and $f$ are certain constants depending on $h$ and its derivatives evaluated at $(u_0,v_0)$ given by
\begin{equation}
 b=\frac{h^0_{uuv}}{h^0_{vv}}, \quad c=\frac{h^0_{uuuu}}{2h^0_{vv}}, \quad d=\frac{h^0_{vv}h^0_{uuuv}-h^0_{uuv}h^0_{uvv}}{(h^0_{vv})^2},
  \quad f=\frac{h^0_{vv}h^0_{uuvv}-2h^0_{vvv}h^0_{uuv}}{2(h^0_{vv})^2}\, . 
\end{equation}
Solving this equation and considering the limit of infinitesimal $\delta u$ and $\delta v$ one gets
\begin{equation}
 v-v_0 \simeq \sqrt{\frac{h^0_{uuuu}}{2h^0_{vv}}}(u-u_0)^2 \, .
\end{equation}
In this case 
\begin{equation}
 \Delta T \simeq (u-u_0)
\end{equation}
which changes sign at the point $u_0$ and, hence, transition is forbidden. \par
In a similar way one can show that in the case when all derivatives
\begin{equation}
 \frac{\partial^k h}{\partial u^k}\Big{\vert}_{h_{uu}=0} \qquad \mathrm{for} \quad k=3,4,5,\dots,n
\end{equation}
the behavior of  near the transition line is of the type
\begin{equation}
 v-v_0\simeq (u-u_0)^{\frac{n+3}{2}}\, .
\end{equation}
So, for odd $n$, the transition is  forbidden allowed while for odd $n$ its is not.\par
Finally if the transition line is given by the equation $h_{vv} =0$ with $h_{uu}|_{h_{vv} =0} \neq 0$ then one has the results
presented above with exchange $u \leftrightarrow v$.\par
For the dNLS equation the transition line $v=0$ is a characteristic and $\frac{\D v }{\D u}\Big|_{v=0}=0$. 
Hence, the transition is not possible. We remark that this case is in some sense degenerate because the Hamiltonian 
is quadratic in $u$ and all the partial derivatives of $\partial^k h/\partial u^n= 0$, $k\geq 3$ are zero. 
For the dB equation, in contrast, the transition line  is $u=0$, characteristic cross
the transition line orthogonally $\frac{\D v }{\D u}\Big|_{u=0}=0$, i.e.
\begin{equation}
 \Delta T= \frac{h_{uuu}}{h_{uuv}}\Big\vert_{u=0} \to \infty,
\end{equation}
and consequently the transition from the hyperbolic domain to elliptic one is not forbidden.
\section{Special classes of Hamiltonian systems: Gas dynamics equations}
\label{sect-exe}
Expressions (\ref{Ham-dNLS}) and (\ref{Ham-dB}) suggest us to consider two special classes of systems with Hamiltonian densities
\begin{equation}
 \begin{split}
  h_1&= F_1(v)+\frac{1}{2}vu^2\, ,\\
  h_2&= F_2(v)+F_3(u)
 \end{split}
\label{exe-H}
\end{equation}
where $F_1,F_2,F_3$ are functions of a single variable\, . 
For the Hamiltonian density of the form $h_1$ one has the system
\begin{equation}
\label{h1-exe}
 \binom{u_t}{v_t}= \left(
 \begin{array}{cc}
  u & {F_1}_{vv} \\ v & u
 \end{array}
\right)  \binom{u_x}{v_x}. 
\end{equation}
and $\Omega=4v{F_1}_{vv}$. Simple waves are defined by equation 
\begin{equation}
 \sqrt{\frac{{F_1}_{vu}}{v}}\D v=\pm \D u\, .
\end{equation}
The relations ${h_1}_{uuu}=0$ and ${h_1}_{uuv}=1$ imply $\Delta T =0$.\par
There are two quite different situations. First corresponds to the case when he transition line is given by $v=0$ and ${F_1}_{vv} >0$ near $v=0$.
For the system  of mixed type one has  
\begin{equation}
 F_1=\frac{1}{2n(n-1)}v^{2n}, \qquad n=1,2,3,\dots\, ,
\end{equation}
the characteristic in $(u,v)$-plane  are given by the equation
\begin{equation}
 \frac{\D v}{\D u}=\pm \sqrt{2n(2n-1)} v^{\frac{3-2n}{2}} 
\end{equation}
and so
\begin{equation}
 \frac{\D^2 v}{\D u^2}\sim v^{2(1-n)}\, .
\end{equation}
Thus if $n \geq 2$ the acceleration $ \frac{\D^2 v}{\D u^2}$ diverges as the characteristic approaches the transition line $v=0$ (which is
also a characteristic) and, hence, the transition is not forbidden. \par
If $v \neq 0$ everywhere
 the properties of the system (\ref{h1-exe}) are quite different.
Introducing the function $P(v)$ defined by the equation
\begin{equation}
 {F_1}_{vv}=\frac{P'(v)}{v},
\end{equation}
one rewrites the system (\ref{h1-exe}) as
\begin{equation}
 \begin{split}
  u_t=&uu_x +\frac{P_x(\rho)}{\rho}\, , \\
  \rho_t=&(u\rho)_x \, .
 \end{split}
 \label{gasdyneqn}
\end{equation}
It is the general isentropic one-dimensional gas-dynamic equation with $u$ being velocity, $v$ being density $\rho$, $P(\rho)$ is the pressure
and $t \to -t$ (see e.g. \cite{Whi,L-VI}). The TL-line is defined by 
\begin{equation}
\label{TL-gas}
\Omega=4 \rho {F_1}_{vv}= 4 P'(\rho)=0 
\end{equation}
and the characteristics in the space $(u,v)$ are defined by the equation
\begin{equation}
 \label{SW-gas}
 \frac{\D \rho}{\D u}= \pm \sqrt{\frac{\rho^2}{P'(\rho)}}\, .
\end{equation}
For ordinary media 
\begin{equation}
  \frac{\D P}{\D \rho}\Big{\vert}_S=c^2\, ,
\end{equation}
i.e. squared sound speed $c$  in the medium. Thus, for normal cases the system is hyperbolic everywhere.  So the system (\ref{gasdyneqn}) is 
of mixed type for particular macroscopic systems for which the derivative $\frac{\D P}{\D \rho}\Big{\vert}_{S}$ can vanish at some value of 
density $\rho_0$ (zero sound speed point) and change sign passing through this value (see e.g. \cite{CHT}-\cite{LQ}). \par
Such a situation is realized, for instance, for the functions $P$ which for small $\rho-\rho_0$ are of the form
\begin{equation}
 \mathrm{a)}: \quad P \sim (\rho-\rho_0)^{2n+2}, \qquad \mathrm{b)}: \quad P \sim (\rho-\rho_0)^\frac{2n+2}{2n+1}\, ,
\end{equation}
where $n=0,\pm 1,\pm 2, \dots$\, . Near the transition line $\rho=\rho_0$ one has
\begin{equation}
 \frac{\D^2 \rho}{\D u^2}\sim (\rho-\rho_0)^{-(1+\gamma)}
\end{equation}
where $\gamma=2n+2$, or $\gamma=\frac{2n+2}{2n+1}$. In both cases $ \frac{\D^2 \rho}{\D u^2} \to \infty $ as $\rho \to \rho_0$ and, hence,
 the transition is not forbidden.
 \section{Nonlinear wave type equations}
 \label{sect-NLW}
The system with Hamiltonian density $h_2$ (\ref{exe-H}) is of the form
\begin{equation}
 \label{NLW-gen}
 \binom{u_t}{v_t}= \left(
 \begin{array}{cc}
  0 & F_2''(v) \\ F_3''(u) & 0
 \end{array}
\right)  \binom{u_x}{v_x}. 
\end{equation}
This system is, in fact, the system of two conservation laws
\begin{equation}
 u_t=(F_2'(v))_x, \qquad v_t=(F_3'(u))_x
\end{equation}
and it is equivalent to the single equation
\begin{equation}
\label{ggNLW}
 (A(w_t))_t=(B(w_x))_x
\end{equation} 
where $w_x=u$, $w_t=F_2'(v)$, $B(y)=F'_3(y)$ and $A(y)= (F_2')^{-1}(y)$. In the particular case $F_2(v)=v^2/2$
equation (\ref{ggNLW}) takes the form 
\begin{equation}
 w_{tt}=(F_3'(w_x))_x
\end{equation}
or 
\begin{equation}
\label{NLW-std}
 u_{tt}=(F_3'(u))_{xx}\, ,
\end{equation}
i.e. the standard form of the nonlinear wave equations (see e.g. \cite{Dub} with $F_3=P$). For the dB equation  
\begin{equation}
F_3^{\mathrm{dB}}=\frac{1}{6}u^3 +\frac{1}{2}c^2 u^2\, .
\end{equation}
For the system (\ref{NLW-gen}) one has
\begin{equation}
\Omega=4F_3''(u)F_2''(v)
\end{equation}
and equations for $(u,v)$ characteristics are given by
\begin{equation}
\label{SW-cond-dB}
 \sqrt{F_2''(v)}\D v = \pm \sqrt{F_3''(u)}\D u \, .
\end{equation}
First we consider the case when the transition line is given by
\begin{equation}
\label{NLW-TL-c1}
 F''_3(u_0)=0, \qquad F''_2(v) \neq 0, 
\end{equation}
which includes the nonlinear wave case (\ref{NLW-std}) where $F_2''(v)=1$. For the system of mixed type near the point $u_0$
the function $F_3(u)$ should be of the form
\begin{equation}
 F_3(u) \sim \it{const} (u-u_0)^{2n+1}, \qquad n=1,2,3,\dots \, . 
\end{equation}
For $(u,v)$ characteristics near the transition line $u=u_0$ one has
\begin{equation}
 \frac{\D v}{\D u} \sim (u-u_0)^\frac{2n-1}{2}\, .
\end{equation}
So $\frac{\D v}{\D u} \to 0$ as $u\to u_0$ and, hence,the characteristic approaches the transition line orthogonally.  Thus the change 
of type is not forbidden. 
It is clearly so for nonlinear wave equations with $F_3(u) = u^{2n+1}$, $n=1,2,3,\dots$ . Note that nonlinear wave equations
with $F_3(u) = u^{2n}$, $n=1,2,3,\dots$ are hyperbolic everywhere. The same is valid for the dispersionless Toda equation $u_{tt}=\exp(u)_{xx}$
for which $F_3(u)=\exp(u)$ (see e.g. \cite{Dub}). \par
If, instead of (\ref{NLW-TL-c1}), the transition line is given by
\begin{equation}
\label{NLW-TL-c2}
 F''_2(u) \neq 0, \qquad F''_3(v_0) = 0, 
\end{equation}
one has the same results with the exchange $u \leftrightarrow v$.
\section{Numerical example of transitions for the dB equation}
\label{sect-num}
Here we present some numerical results for the dB equation as the characteristic representative of the generic class of Hamiltonian systems. \par
\begin{figure}
\centering
{\includegraphics[width=10cm]{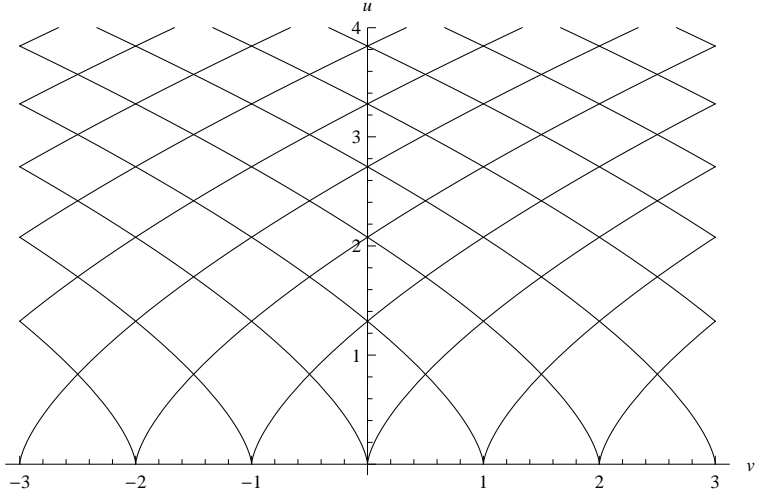}}
\caption{Simple wave  structure (\ref{SWdB}) for dB equation (\ref{dBi}).} 
\label{SWdB-fig}
\end{figure}

Let us consider the class of periodic solutions with fixed boundary values and with initial conditions
\begin{equation}
\label{PbCexe}
 (u(x,0)-c)^2+v^2=1, \qquad u(x,0)=c+\sin(x) 
\end{equation}
where $c>1$ is assumed in order to start in the hyperbolic sector.
The simple waves for dB equation are (see figure \ref{SWdB-fig})
\begin{equation}
\label{SWdB}
 v \pm \frac{2}{3}u^{3/2} =  k, \qquad k \in \mathbb{R} \, .
\end{equation}
For every value of $c$ there are four simple waves tangent to the circle. The two lower ones 
satisfy the system
\begin{equation}
\begin{split}
 \left(\frac{3}{2}(k \pm v_c)\right)^{2/3}= c-\sqrt{1-v_c^2} \, , \\
 \frac{1}{(k-v_c)^{1/3}}=\frac{v_c}{\sqrt{1-v_c^2}}\, .
\end{split}
\label{vtcond}
\end{equation}
where $v_c$ is the value of $v$ at contact point. 
For every value of $c>1$ the previous systems admits two solutions corresponding to two different simple waves symmetric with respect to the $u$
axis. Only above a minimum value of $c$, called $c_{crit}$, the couple of tangent simple waves 
intersects each other. Therefore only above $c_{crit}$ the hyperbolic elliptic transition is forbidden because the initial condition 
(see e.g. \cite{T6}) is separated from the transition line by the two tangent simple waves. \par
The critical value $c_{crit}$ can be estimated as follows. 
At $c_{crit}$ the intersection of the tangent simple waves, because of the problem symmetry, is at the origin $u=0$, $v=0$, i.e. with 
$k=0$ in  (\ref{vtcond}).  Solving therefore (\ref{vtcond}) with $k=0$ we obtain the
value of this critical constant (in case (\ref{PbCexe})) which is $c_{crit} \simeq 1.7472$.\par
\begin{figure}
\centering
{\includegraphics[width=10cm]{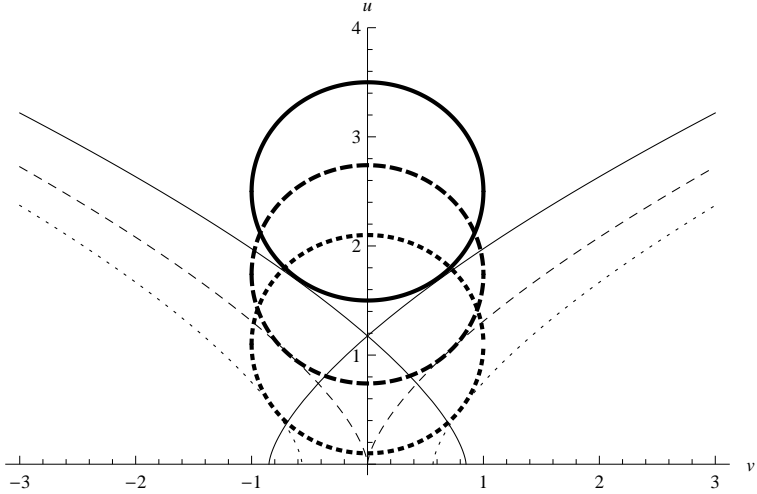}}
\caption{In the example of initial conditions class (\ref{PbCexe}) we shown that, in dependence of the parameter values the hyperbolic-elliptic 
transition is forbidden (solid line) or allowed (dotted line).} 
\label{dB-circle-transition-fig}
\end{figure}
\begin{figure}
\centering
\begin{tabular}{cccc}
\includegraphics[width=7cm]{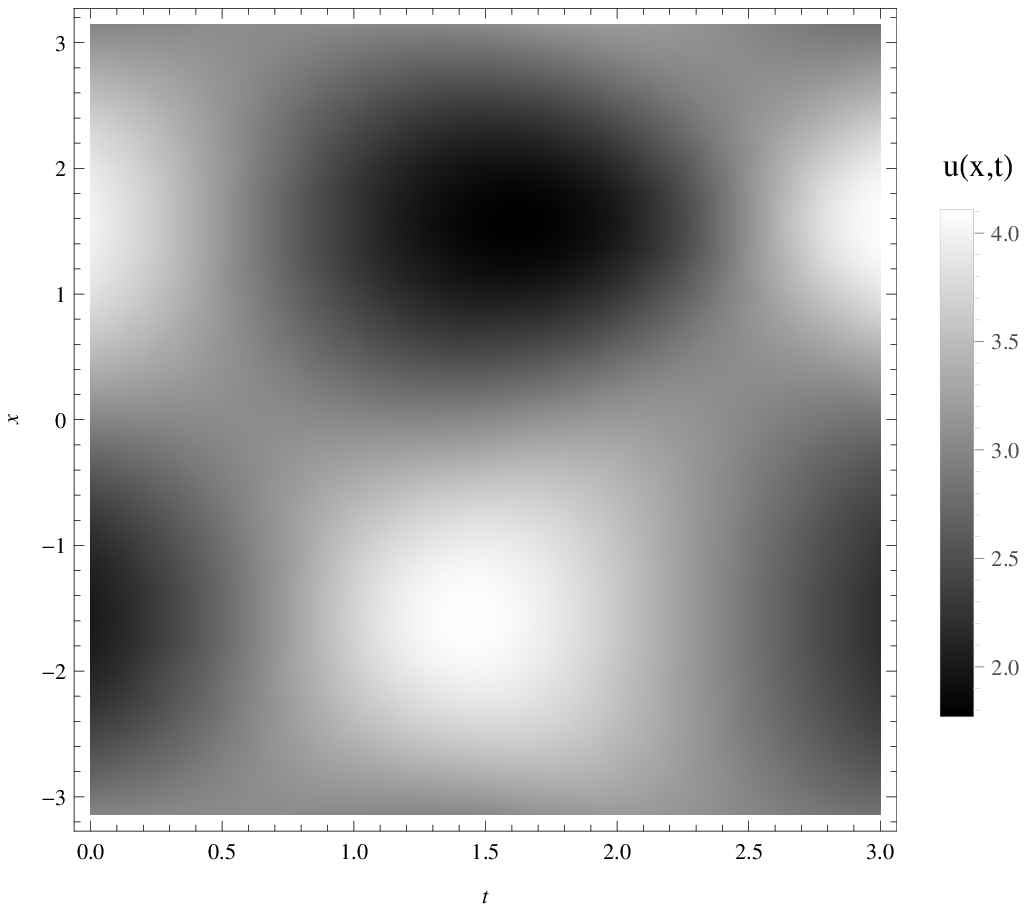} & \phantom{a} \hspace{2cm} & 
\includegraphics[width=7cm]{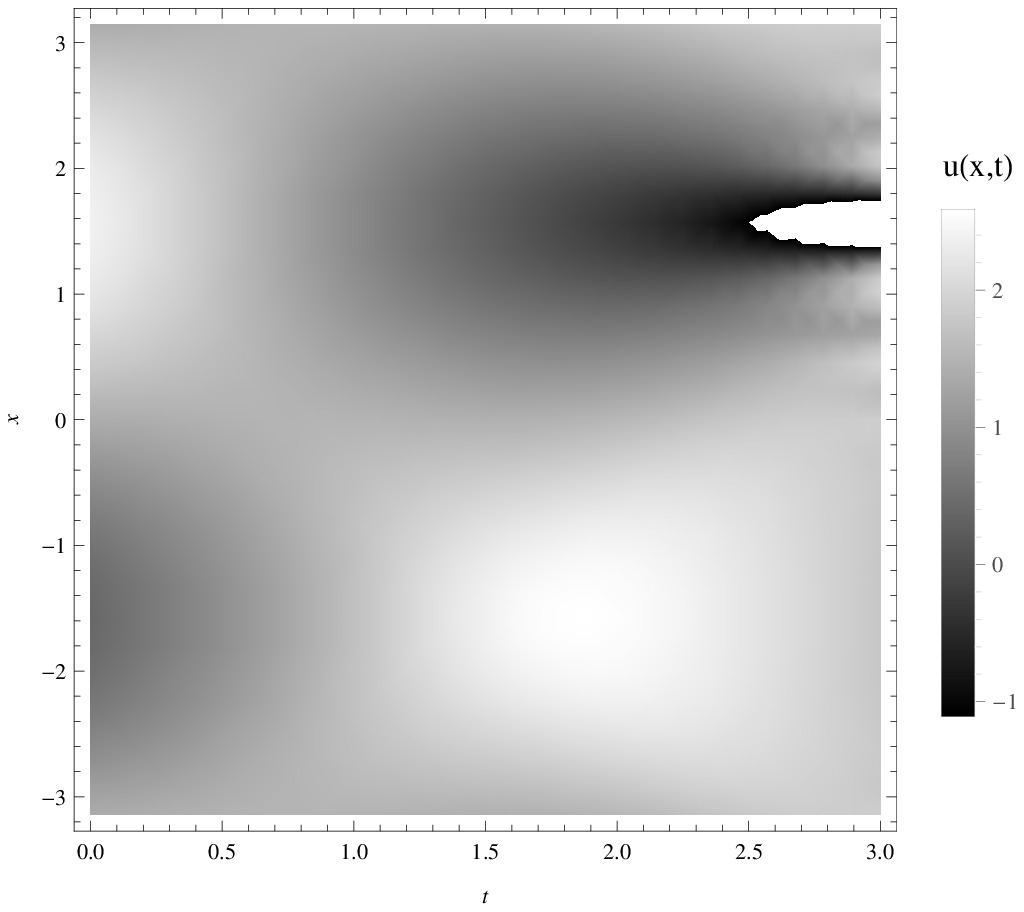}
\end{tabular}
\caption{On the right plot  we show the evolution of $u$ with $c=3$ in the initial conditions (\ref{PbCexe}) while on the left plot with $c=1.4$. 
The motion remains in the hyperbolic sector if $c=3>c_{crit}$, while in the case $c=1.4<c_{crit}$ we see the transition in the 
elliptic sector (upper-right region of the plot).} 
\label{u-per}
\end{figure}
\begin{figure}
\centering
\includegraphics[width=10cm]{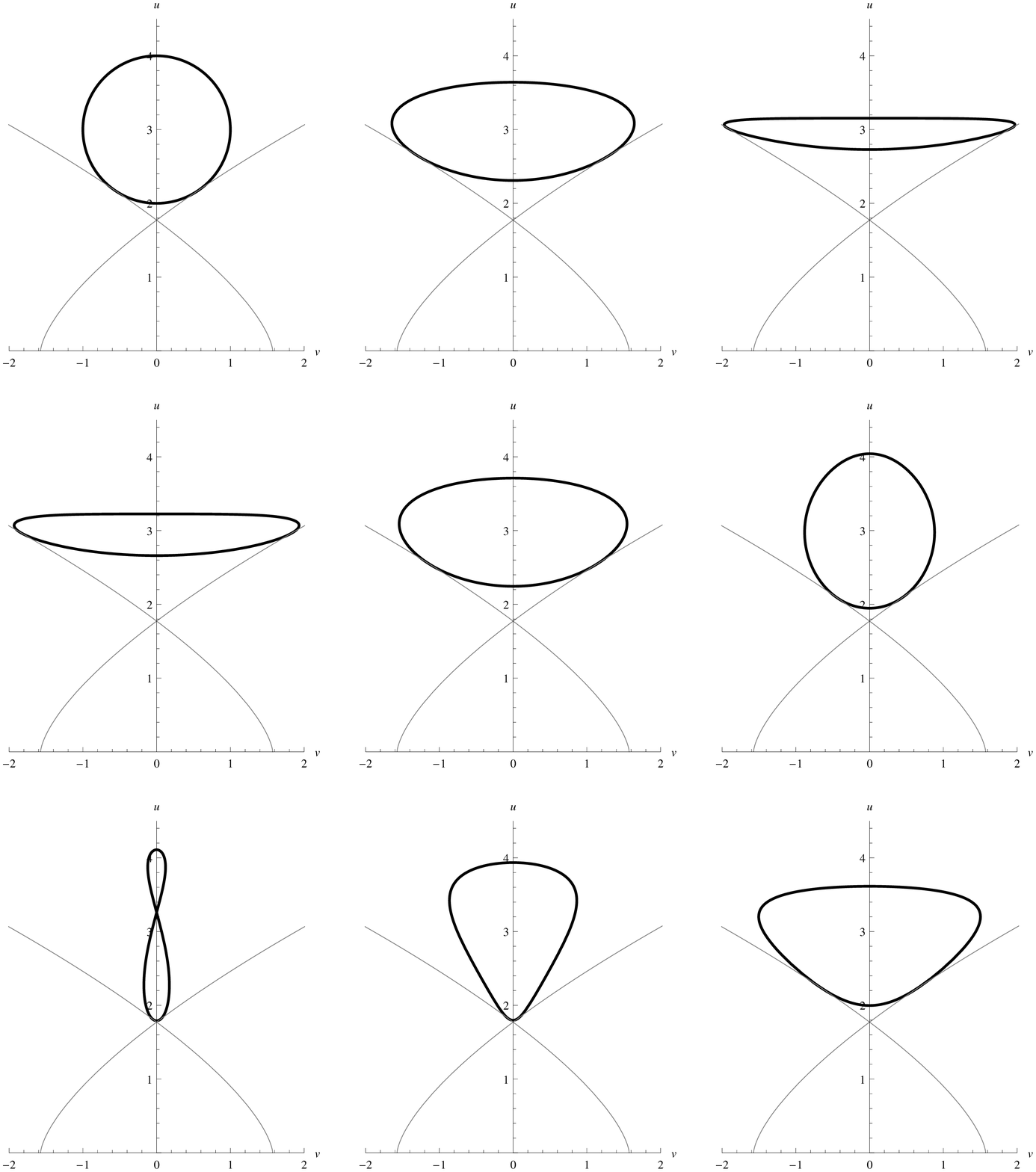}
\caption{Circle hyperbolic evolution with $c=3> c_{crit}$, $t$ from zero to $t=2$ with equispaced steps.
The curve evolution does not admits the transition below the intersection point.}
\label{circle-hyp}
\end{figure}
\begin{figure}[!ht]
\centering
\includegraphics[width=10cm]{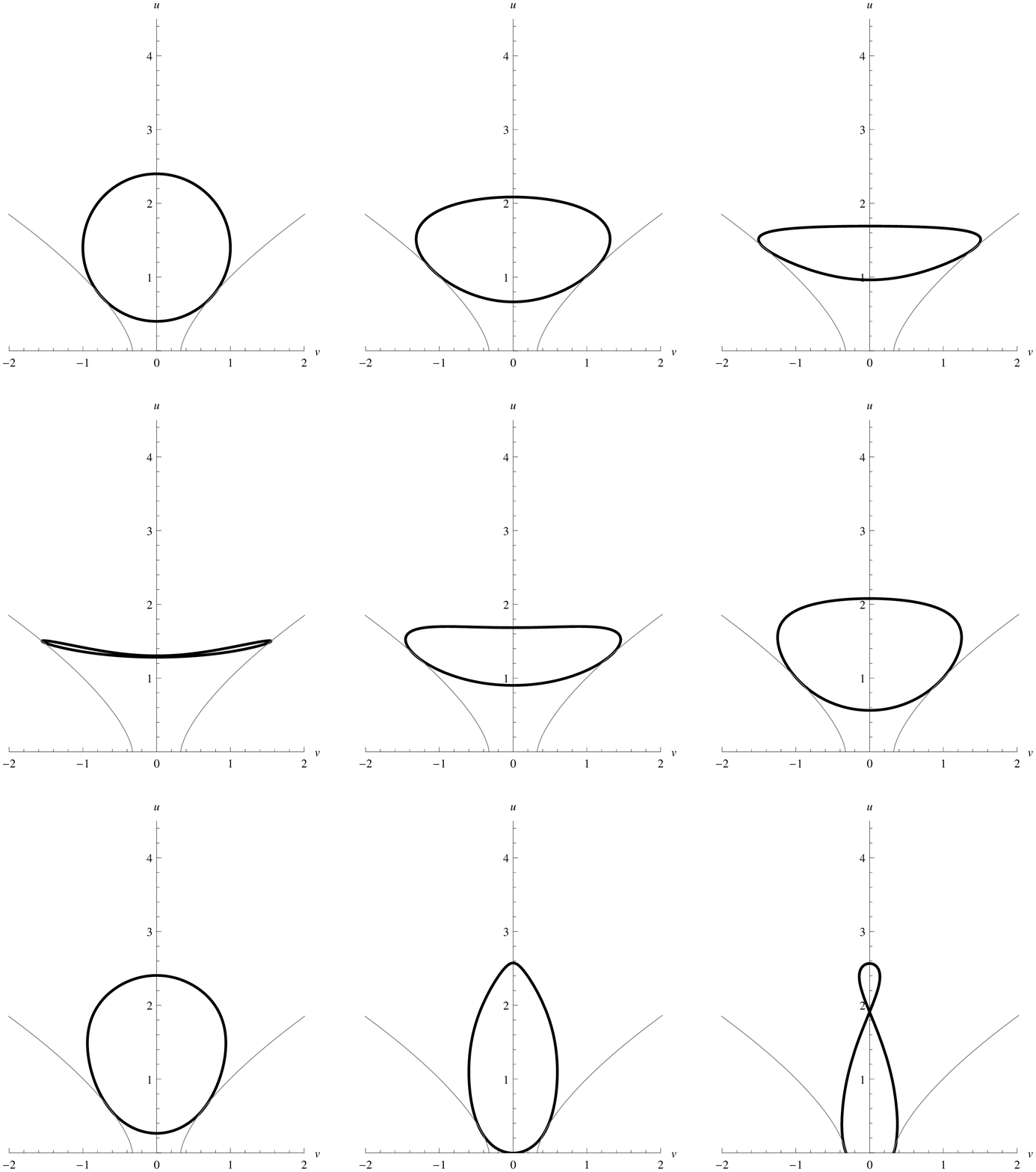}
\caption{Circle  evolution with $c=1.4<c_{crit}$, $t$ from zero to $t=2$ with equispaced steps.  The curve  can cross the 
transition line $u=0$.}
\label{circle-ell}
\end{figure}

In figure \ref{dB-circle-transition-fig}  the hyperbolic elliptic transition possibility 
as the function of the parameter $c$ of the family of initial conditions (\ref{PbCexe}) is shown.  
The solid circle are the initial data at $c=3$ which is greater than the critical value. In this case the
initial data are bounded from below by two simple waves (two solid open curves) which are tangent to the initial data and intersect each other above 
the transition line $u=0$. This behavior  prevents the transition. The dashed circle (initial conditions with 
$c = c_{crit}$) are the critical conditions in the family which forbid the transition. Actually the two tangent simple waves  
(dashed open curves) intersect exactly at the transition line. Finally the dotted circle are initial data at $c=1.4$ which is below the 
critical value: the tangent simple waves (two dotted curves) have no intersection and the initial data could reach  
the transition line. 
In figures \ref{u-per} (left) and \ref{circle-hyp} it is shown the evolution of $u$ in dependence of $(x,t)$ and $v$ 
respectively in the non transition case of $c=3>c_{crit}$. 
In figures \ref{u-per} (right) and \ref{circle-ell} is shown the evolution of $u$ in dependence of $(x,t)$ and $v$ 
respectively in the transition case of $c=1.4<c_{crit}$. At the second to last step the curve is tangent 
to the transition line. However this line is not a characteristic in the $(u,v)$ space and the curve can cross the transition line as can
be seen in the last plot.\par
The numerical evolutions in the figures \ref{u-per}, \ref{circle-hyp} and \ref{circle-ell}
are obtained using Mathematica.
\subsubsection*{Acknowledgments}
 G.O. thanks R. Camassa for useful discussions and informations.  
 Partial support by MIUR Cofin 2010-2011 project 2010JJ4KPA is
acknowledged. 
 This work was carried out under the auspices of the GNFM Section
of INdAM. \newpage 

\end{document}